\documentclass[useAMS,usenatbib]{mn2e}
\usepackage{graphicx}
\usepackage{latexsym}
\usepackage{amsmath}
\usepackage{amssymb}

\title[Are group and cluster halos over-concentrated?]{Are group- and cluster-scale dark matter halos over-concentrated?}

\author[M.~W.~Auger et al.]{M.~W.~Auger$^1$, J.~M.~Budzynski$^1$, V.~Belokurov$^1$, S.~E.~Koposov$^{1,2}$, I.~McCarthy$^{1,3,4}$\\
$^1$ Institute of Astronomy, Madingley Road, Cambridge, CB3 0HA, UK\\
$^2$ Sternberg Astronomical Institute, Moscow MV Lomonosov State University, Moscow 119992, Russia\\
$^3$ School of Physics and Astronomy, University of Birmingham, Edgbaston, Birmingham B15 2TT, UK\\
$^4$ Astrophysics Research Institute, Liverpool John Moores University, Birkenhead CH41 1LD, UK}

\begin{document}

\pagerange{\pageref{firstpage}--\pageref{lastpage}} \pubyear{2002}

\newcommand{\dd}{\textrm{d}}
\newcommand{\rin}{R_\textrm{\tiny{in}}}
\newcommand{\OO}{\mathcal{O}}

\maketitle

\label{firstpage}

\def\mnras{MNRAS}
\def\apj{ApJ}
\def\aap{A\&A}
\def\aapr{A\&A Reviews}
\def\aaps{A\&A}
\def\apjl{ApJL}
\def\aj{AJ}
\def\apjs{ApJS}
\def\araa{ARA\&A}
\def\nat{Nature}
\def\pasp{PASP}


\begin{abstract}
We investigate the relationship between the halo mass, $M_{200}$, and concentration, $c$, for a sample of 26 group- and cluster-scale strong gravitational lenses. In contrast with previous results, we find that these systems are only $\sim 0.1$ dex more over-concentrated than similar-mass halos from dark matter simulations; the concentration of a halo with $M_{200} = 10^{14} M_\odot$ is log~$c = 0.78\pm0.05$, while simulations of halos with this mass at similar redshifts ($z \sim 0.4$) predict log~$c \sim 0.56 - 0.71$. We also find that we are unable to make informative inference on the slope of the $M_{200}-c$ relation in spite of our large sample size; we note that the steep slopes found in previous studies tend to follow the slope in the covariance between $M_{200}$ and $c$, indicating that these results may be measuring the scatter in the data rather than the intrinsic signal. Furthermore, we conclude that our inability to constrain the $M_{200}-c$ slope is due to a limited range of halo masses, as determined by explicitly modelling our halo mass distribution, and we suggest that other studies may be producing biased results by using an incorrect distribution for their halo masses.
\end{abstract}



\section{Introduction}\label{sec:intro}
Dark matter only simulations predict a tight power law relationship between the masses and concentrations of dark matter halos that has a logarithmic slope of $\sim -0.1$ \citep[e.g.,][]{bullock,neto,maccio,duffy}. The normalisation of this trend agrees with observations of massive galaxy clusters \citep[$M \sim 10^{15}$; e.g.,][and references therein]{ettori}, but lower-mass clusters and rich groups appear to deviate from the predicted relation by nearly an order of magnitude \citep[e.g.,][]{oguri12}. This tension between observations and simulations may be due in part to selection effects, as mass measurements of groups and poor clusters typically rely on strong gravitational lensing. More concentrated halos are more efficient lenses, and strong lensing may also preferentially occur in halos that are oriented such that the long axis of the mass distribution is along the line of sight \citep[e.g.,][]{hennawi}, an effect that can mimic higher concentrations. However, both of these biases are weak, and \citet{oguri12} explicitly include these effects in their analysis but still find that the observed mass-concentration trend is significantly stronger than the relation in simulations.

The paucity of well-studied groups and low-mass clusters makes it difficult to characterise the mass-concentration relation below $M \sim 10^{14} M_\odot$, where the tension with theory is most significant. To remedy this, we have compiled a sample of $\sim$100 group-scale gravitational lens systems identified by the CASSOWARY survey \citep{cswa2}. The strong lensing signals can be combined with total masses derived from group richnesses \citep[e.g.,][]{wiesner} to determine halo concentrations for these systems with masses potentially as low as $M\approx10^{13} M_\odot$. Here we present a preliminary analysis of 26 lenses, and we find that our objects are only slightly more concentrated than simulations predict. We introduce the sample and our mass measurements in Section 2, describe our analysis techniques in Sections 3 and 4, then compare with previous studies in Section 5. We assume a $\Lambda$CDM cosmology with $\Omega_\Lambda = 0.7$ and $h = 0.7$.

\section{Lens Sample and Mass Measurements}\label{sec:data}

The CASSOWARY sample includes nearly 100 strong gravitational lens candidates discovered in Sloan Digital Sky Survey \citep[SDSS;][]{SDSSyork} imaging data \citep[e.g.,][]{cswa1,cswa2}. The selection algorithm searches around luminous red galaxies (LRGs) for resolved blue neighbours that potentially correspond to $z \sim 1-3$ gravitationally lensed arcs. The arcs are required to be bright and the distances between the arcs and the LRG must lie in the range 1.5\arcsec to 15\arcsec. This imaging selection preferentially selects Einstein radii between galaxy-scale lenses \citep[e.g., SLACS;][]{bolton06} and massive cluster-scale lenses \citep[e.g.,][]{oguri12}, enabling our investigation of group-scale dark matter halos. In this paper we focus on 26 lenses that have sufficiently good SDSS imaging to robustly determine Einstein radii. Additionally we require that all of the lenses have well-constrained lens and source redshifts. All of the source redshifts and 23 of the lens redshifts are spectroscopically determined \citep{stark}, whilst three of the lenses have photometric redshifts;  the mean lens redshift of the sample is $z = 0.4$, and details for the lenses can be found in Table \ref{T_lenses}.

\begin{table*}
 \begin{center}
 \caption{Properties of the 26 Lenses. Columns: (1) lens name; (2) right ascension (J2000); (3) declination; (4) lens redshift; (5) source redshift; (6) Einstein radius in arcseconds; (7) richness estimator; (8) uncertainty on the richness; (9) inferred mass within $r_{500}$; (10) uncertainty on log~$M_{500}$; (11) $r$-band luminosity in solar units. Daggers indicate the Einstein radii for which we have adopted a 25\% uncertainty, while the other systems have 5\% uncertainty.}
 \begin{tabular}{@{}lrrcclrcccc@{}}
 \hline
 Name  &  RA  &  Dec  &  $z_{\rm lens}$  &  $z_{\rm src}$  &  $r_{\rm Ein} (^{\prime\prime})$  &  N$_{\rm 1Mpc}$  &  $\sigma_{N_{\rm 1Mpc}}$  &  log $M_{500}$  &  $\sigma_{M_{500}}$  &  log $L_r$ \\
 \hline
 SDSSJ0022+1431  &    5.6704915  &   14.5195652  &  0.3800  &  2.7300  &              3.26  &  11.987  &  3.101  &  14.277  &  0.245  &  11.39 \\
SDSSJ0105+0144  &   16.3318837  &    1.7489944  &  0.3613  &  2.1300  &              3.48  &   5.213  &  3.489  &  13.876  &  0.411  &  11.36 \\
SDSSJ0143+1607  &   25.9588685  &   16.1274983  &  0.4360  &  1.5100  &              2.67  &   0.413  &  2.427  &  13.358  &  0.539  &  11.42 \\
SDSSJ0145-0455  &   26.2678898  &   -4.9309974  &  0.6040  &  1.9580  &              1.93  &   0.880  &  1.010  &  13.159  &  0.486  &  11.45 \\
SDSSJ0146-0929  &   26.7333707  &   -9.4979150  &  0.4400  &  1.9440  &             11.94  &  24.240  &  2.682  &  14.602  &  0.217  &  11.72 \\
SDSSJ0232-0323  &   38.2077827  &   -3.3905627  &  0.4500  &  2.5180  &              3.66  &   2.760  &  2.579  &  13.631  &  0.457  &  11.67 \\
SDSSJ0807+4410  &  121.8812994  &   44.1801468  &  0.4490  &  2.5360  &  2.06$^{\dagger}$  &   4.053  &  2.163  &  13.767  &  0.370  &  11.23 \\
SDSSJ0846+0446  &  131.6977631  &    4.7680718  &  0.2410  &  1.4052  &  3.36$^{\dagger}$  &  17.667  &  3.601  &  14.454  &  0.229  &  11.24 \\
SDSSJ0854+1008  &  133.6196933  &   10.1374270  &  0.2980  &  1.4370  &  4.16$^{\dagger}$  &   8.507  &  2.621  &  14.119  &  0.265  &  11.28 \\
SDSSJ0900+2234  &  135.0110189  &   22.5680156  &  0.4890  &  2.0325  &              7.90  &   8.640  &  2.331  &  14.132  &  0.247  &  11.35 \\
SDSSJ0952+3434  &  148.1676045  &   34.5794650  &  0.3490  &  2.1896  &  4.16$^{\dagger}$  &  22.253  &  5.254  &  14.550  &  0.237  &  11.03 \\
SDSSJ0957+0509  &  149.4132999  &    5.1588678  &  0.4400  &  1.8230  &  5.39$^{\dagger}$  &   1.853  &  1.569  &  13.440  &  0.451  &  11.24 \\
SDSSJ1115+1645  &  168.7683082  &   16.7607213  &  0.6030  &  1.7180  &  4.57$^{\dagger}$  &   3.307  &  0.735  &  13.720  &  0.243  &  11.39 \\
SDSSJ1137+4936  &  174.4169025  &   49.6098718  &  0.4480  &  1.4110  &  2.80$^{\dagger}$  &   4.040  &  2.499  &  13.763  &  0.405  &  11.26 \\
SDSSJ1138+2754  &  174.5373066  &   27.9085314  &  0.4470  &  0.9090  &  6.15$^{\dagger}$  &  23.267  &  3.255  &  14.577  &  0.219  &  11.42 \\
SDSSJ1147+3331  &  176.8470946  &   33.5315538  &  0.2120  &  1.2050  &  4.58$^{\dagger}$  &  17.307  &  2.628  &  14.448  &  0.220  &  11.27 \\
SDSSJ1148+1930  &  177.1380745  &   19.5008726  &  0.4440  &  2.3790  &              5.12  &   1.333  &  3.391  &  13.568  &  0.503  &  11.45 \\
SDSSJ1206+5142  &  181.5087122  &   51.7082044  &  0.4330  &  2.0000  &              3.88  &   5.867  &  2.040  &  13.949  &  0.287  &  11.55 \\
SDSSJ1209+2640  &  182.3486973  &   26.6796452  &  0.5580  &  1.0180  &  8.44$^{\dagger}$  &   9.960  &  2.254  &  14.202  &  0.236  &  11.79 \\
SDSSJ1240+4509  &  190.1345157  &   45.1507908  &  0.2740  &  0.7252  &  2.93$^{\dagger}$  &   8.680  &  2.436  &  14.132  &  0.250  &  10.97 \\
SDSSJ1450+3908  &  222.6276986  &   39.1386469  &  0.2890  &  0.8613  &  3.39$^{\dagger}$  &   8.307  &  3.251  &  14.094  &  0.310  &  11.34 \\
SDSSJ1511+4713  &  227.8280648  &   47.2278725  &  0.4520  &  0.9800  &  4.40$^{\dagger}$  &   5.907  &  1.986  &  13.957  &  0.278  &  11.73 \\
SDSSJ1629+3528  &  247.4773470  &   35.4776361  &  0.1700  &  0.8500  &  3.58$^{\dagger}$  &  12.253  &  3.258  &  14.284  &  0.248  &  10.95 \\
SDSSJ1958+5950  &  299.6471715  &   59.8496884  &  0.1800  &  2.2200  &  6.19$^{\dagger}$  &   6.040  &  4.324  &  13.944  &  0.415  &  11.53 \\
SDSSJ2158+0257  &  329.6819958  &    2.9583904  &  0.2850  &  2.0800  &              3.47  &   8.627  &  2.668  &  14.129  &  0.263  &  11.33 \\
SDSSJ2222+2745  &  335.5357320  &   27.7598969  &  0.4850  &  2.8070  &              7.96  &   9.040  &  2.487  &  14.150  &  0.252  &  11.62 \\

 \hline
 \label{T_lenses}
 \end{tabular}
 \end{center}
\end{table*}

\subsection{Strong Lensing}\label{sub:masslens}

Strong lensing provides a precise and robust measurement of the Einstein radius, i.e., the radius within which the mean surface mass density is equal to the lensing critical density. We use the SDSS imaging data to constrain the Einstein radii for each of our 26 lenses, and the mass distributions of the lenses are assumed to be singular isothermal ellipsoids; although this model may be a poor description of the true density profile, the inferred Einstein radii are nevertheless robust \citep[e.g.,][]{skw06}. Following the modelling scheme of \citet{auger11}, we use an optimiser to simultaneously find the best mass and surface brightness models for the lens and the source in the $g$, $r$, and $i$ imaging from SDSS. The Einstein radius is typically constrained at the 1-2\% level but we impose a minimum uncertainty of 5\% to account for systematics, including line-of-sight structure. In some cases there is no visible counter-image and the Einstein radius becomes significantly degenerate with the flattening of the mass distribution, and for these lenses we therefore impose a conservative 25\% uncertainty on the Einstein radii \citep[compare with the galaxy-scale lenses that lack counter images in][]{brewer}; this is comparable to the spread of reasonable models for the least-constrained lens systems.

\subsection{Mass estimate from optical richness}\label{sub:massopt}

We supplement the central mass constraint from lensing with an estimate of $M_{500}$ based upon the number of group/cluster galaxies around the lens. Our richness estimates are defined for 1 Mpc apertures as in \citet{jbud}, but here we employ some slight modifications. We select objects from the SDSS Data Release 9 catalogue within 5 Mpc of the lens that meet the following criteria: absolute magnitude brighter than $M_{\rm r} = -21.2$, photometric uncertainties less than 0.2 mag in the $griz$ filters, and photometric redshifts within 5\% of the lens redshift. These requirements lead to a small bias towards under-richness for the highest-redshift systems ($z \gtrsim 0.6$) due to group members not being detected in the bluer filters, but loosening these criteria leads to a much noisier estimator and the uncertainties for the highest redshift systems are larger than the bias.

We determine the `background' density of objects in an annulus between 2.5 Mpc and 5 Mpc and subtract the inferred number of background objects from the number of objects in the central 1 Mpc to determine our richness statistic. The uncertainty on the richness is dominated by the variance of the cosmological background structure, and we estimate this by determining the variance between random 1 Mpc radius patches in the annulus between 2.5 and 5 Mpc. We use the same sample of X-ray luminous clusters from \citet{jbud} to calibrate the mass-richness relation, and we find that our richness statistic produces a relation that has a slope of $1.00\pm0.07$, intercept $14.21\pm0.03$, and scatter $0.21\pm0.02$. Typical uncertainties on our $M_{500}$ estimates range from 0.2 dex for the most rich systems to 0.5 dex for the sparsest lenses.

\section{Group and Cluster Mass Models}

We model our lenses with two mass components, one describing the dark matter distribution and the other describing the baryonic mass distribution. The dark matter distribution is modelled as an NFW profile with two free parameters, the mass $M_{200}$ and the concentration $c$ \citep[there is some evidence that real dark matter halos may deviate from NFW halos within the scale radius, e.g.,][but here we constrain our analysis to an NFW context]{newman,sonnenfeld,grillo}. The baryonic mass is only significant for the strong lensing, and we therefore approximate the total baryonic mass with the mass of the central lensing galaxy (note that the baryonic contribution to $M_{500}$ is \textit{at most} the cosmological fraction, i.e., 0.07 dex, which is much smaller than the uncertainties on $M_{500}$). We use the $r-$band luminosity inferred from the SDSS photometry and assume a broad uniform prior on the mass-to-light ratio $\Upsilon$ between 1.8 and 3.2 (i.e., spanning a range between Chabrier and Salpeter initial mass functions). Our model therefore has three free parameters for each lens: $M_{200}$, $c$, and $\Upsilon$.

Our model mass distributions are treated as spherical, and it is therefore straightforward to calculate the observables, $r_{\rm Ein}$ and $M_{500}$, from the models (Appendix A). The normalisations of the dark matter halo and the bulge are perfectly degenerate for a single mass measurement, but the degeneracy can be broken with multiple aperture mass measurements if the halo profile and the light profile are sufficiently different. However, these mass normalisations are still degenerate with the scale (i.e., concentration) of the dark matter halo if only two aperture mass estimates are available, although the model can be made tractable if we assume that there is some underlying physics or phenomenology that relates at least two of the model parameters. Fortunately, the mass-concentration relation provides the phenomenology that helps to break this remaining degeneracy.

\section{Analysis}\label{sec:analysis}

We use the data described in Section 2 to constrain the properties of the model described in Section 3. The inference on the model parameters is given by employing Bayes' theorem,
\begin{eqnarray}
\nonumber \mathrm{P}(c,M_{200},\Upsilon|\widetilde r_{\rm Ein},\widetilde M_{\rm 500}) & = & \mathrm{P}(\widetilde r_{\rm Ein},\widetilde M_{\rm 500}|c,M_{200},\Upsilon)  \\
& & \times \frac{\mathrm{P}(c,M_{200},\Upsilon)}{\mathrm{P}(\widetilde r_{\rm Ein},\widetilde M_{\rm 500})},
\label{eqn_bayesprob}
\end{eqnarray}
where tildes describe observed quantities and
\begin{equation}
\mathrm{P}(\widetilde r_{\rm Ein},\widetilde M_{\rm 500}|c,M_{200},\Upsilon) = \mathrm{P}(\widetilde r_{\rm Ein}|r_{\rm Ein})\mathrm{P}(\widetilde M_{500}|M_{500})
\label{eqn_likelihood}
\end{equation}
with $r_{\rm Ein}$ and $M_{500}$ functions of $c$, $M_{200}$, and $\Upsilon$ as described in Appendix A. The two probabilities on the righthand side of Equation \ref{eqn_likelihood} are given by normal (i.e., Gaussian) distributions, e.g.,
$$
\mathrm{P}(\widetilde r_{\rm Ein}|r_{\rm Ein}) \sim \mathrm{N}(r_{\rm Ein}(c,M_{500},\Upsilon),\sigma^2_{\rm Ein}),
$$
where $\sigma_{\rm Ein}$ is the measurement uncertainty on $\widetilde r_{\rm Ein}$.

Our inference on the mass and concentration of a typical object is shown in Figure 1. We have assumed a uniform prior of $1.8 < \Upsilon < 3.2$ and subsequently marginalised over the mass-to-light ratio. The requirement for strong lensing excludes low-mass, low-concentration halos (the grey region), and we show the independent constraints from using only the richness (red contours), only the Einstein radius (blue contours), and the joint constraint from both (closed black contours). We note that none of these distributions are well-approximated by Gaussians, and in our evaluation of the mass-concentration relation we therefore use this full likelihood space.

\begin{figure}
\centering
\includegraphics[width=0.49\textwidth,clip]{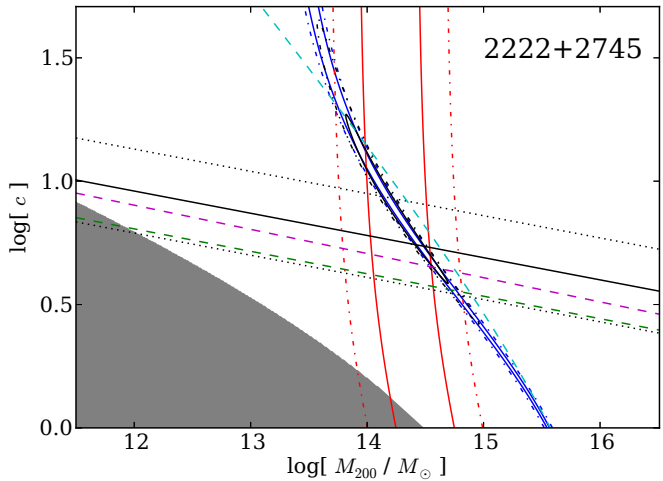}
\caption{The inference on the halo mass and concentration for one lens, SDSSJ2222+2745, from strong lensing (blue contours), richness (red contours), and both constraints (black contours); the solid and dot-dashed contours define the regions that contain 68\% and 95\% of the probability, respectively. The gray-shaded area in the lower left corner is excluded by our lensing selection (these halos would not yield Einstein radii greater than 1\arcsec for all of the $\Upsilon$ values investigated). The black solid line is our best-fit mass-concentration relation, including a strong prior on the slope, with the 1-$\sigma$ scatter shown with black dotted lines. The green and magenta dashed lines are simulations-derived relations from \citet{duffy} and \citet{maccio} respectively, and the cyan dashed line is from the observations of \citet{oguri12}.}
\label{fig_inference}
\end{figure}

\subsection{The Mass-Concentration Relation}
The numerator of the final term in Equation 1, $\mathrm{P}(c,M_{200},\Upsilon)$, is formally the \textit{prior} on our mass model parameters $c$, $M_{200}$, and $\Upsilon$. However, the mass-concentration relation from simulations \citep[e.g.,][]{bullock,neto,maccio,duffy} acts as a conditional prior on the concentration, so that \begin{equation}
\mathrm{P}(c,M_{200},\Upsilon) = \mathrm{P}(c|M_{200}){\rm P}(M_{200}){\rm P}(\Upsilon)
\end{equation}
where
\begin{equation}
\label{eqn_massConc}
\mathrm{P}(\mathrm{log}\ c|{\rm log}\ M_{200}) \sim \mathrm{N}(a~\mathrm{log}\ M_{200}/10^{14} + b,\sigma^2).
\end{equation}
Therefore, fitting for the mass-concentration relation (i.e., determining $a$, $b$, and $\sigma$ in Equation 4) is an example of \textit{hierarchical inference}. Indeed, our problem is the same as the `fitting a straight line with scatter to noisy data' problem addressed by \citet{kelly}, with the exception that our `data' -- here, our inference on the mass and concentration of each lens, as shown by the black contours of Figure 1 -- have highly-correlated \textit{non-Gaussian} uncertainties.

The prior on the halo mass, P(${\rm log}\ M_{200}$), should, in principle, be chosen to encode the parent distribution of our sample of groups, including the halo mass function and the selection imposed by lensing (i.e., that more concentrated and/or more massive halos have a larger cross-section for strongly lensing a background source). However, the CASSOWARY selection function is extremely difficult to determine from first principles, and we therefore choose to also \textit{infer} the underlying distribution from which our lenses are sampled. \citet{kelly} also addresses this problem, and he develops a framework that uses a mixture of Gaussian components to describe the underlying distribution of the independent variable (here, ${\rm log}\ M_{200}$); in this work, for the sake of simplicity we choose to investigate two different models for P(${\rm log}\ M_{200}$). The first model assumes that our masses are drawn from a uniform distribution with central mass log $M_{\rm U}$ and half-width $\Delta M_{\rm U}$, while the second model assumes that our masses are drawn from a normal distribution with mean log $M_{\rm N}$ and variance $\sigma^2_{\rm N}$ (i.e., a Gaussian mixture model with a single component). The priors on log $M_{\rm U}$ and log $M_{\rm N}$ are normal distributions centred on 14 with variance 1 dex, whilst the priors on the widths are both uniform in the log of $\Delta M_{\rm U}$ and $\sigma_{\rm N}$ over several decades from 0.02 to 2 dex.

A comprehensive description of the statistical model we use to turn our measurements of $\widetilde r_{\rm Ein}$ and $\widetilde M_{500}$ in to inference on $a$, $b$, and $\sigma$ is provided in Appendix B.

\subsection{Results}
\begin{figure*}
\centering
\includegraphics[width=0.96\textwidth,clip]{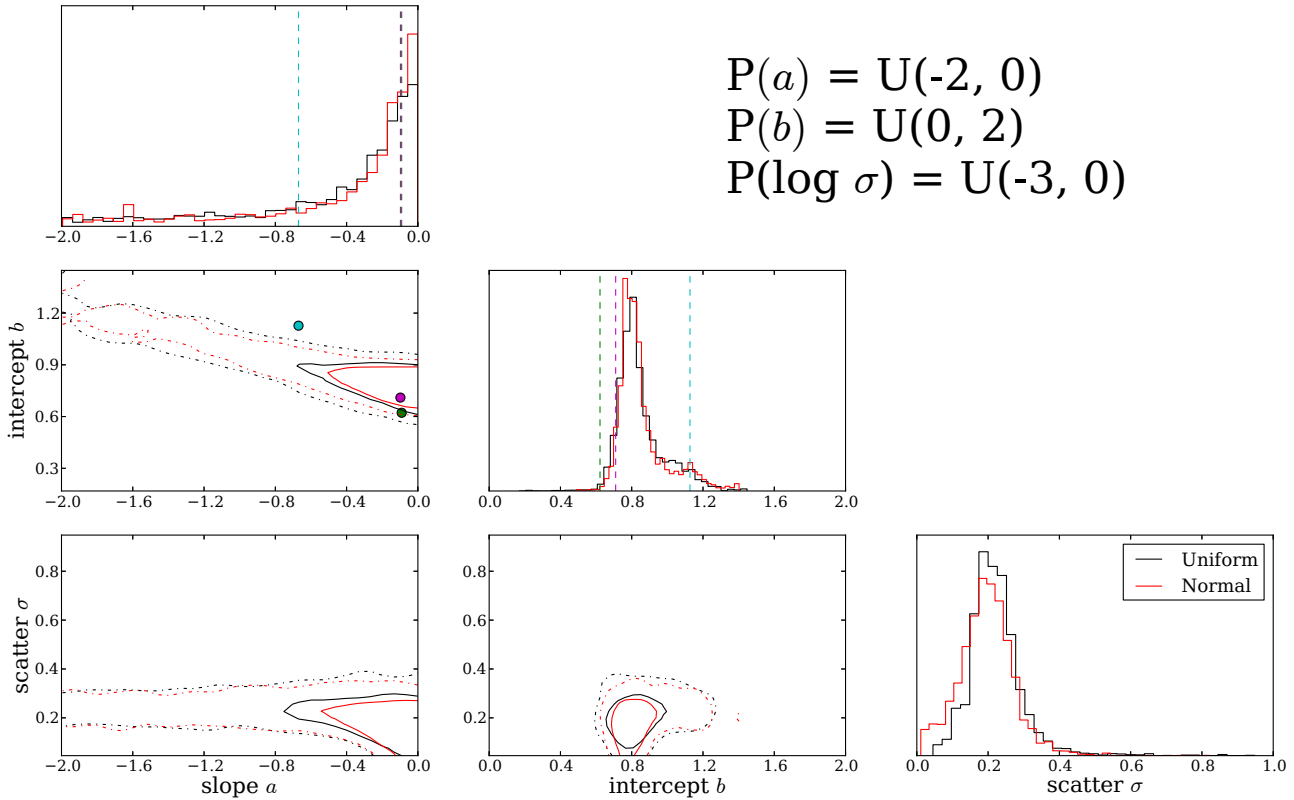}
\caption{Inference on the mass-concentration relation parameters for the CASSOWARY lenses, assuming a broad uniform prior on the slope; solid contours and dot-dashed contours enclose 68\% and 95\% of the marginalised posterior probability, respectively. The black contours are for our model that assumes the halo masses are drawn from a uniform distribution and the red contours are the results when we assume a normal distribution. We find that the lenses prefer a shallow mass-concentration relation with a very extended tail to steep slopes. The intercept implies a concentration of log~$c = 0.85\pm0.13$ (i.e., $c \approx 7.1\pm2$) for a halo with mass $M_{200} = 10^{14} M_\odot$; by comparison, \citet{oguri12} find a concentration of 14 from strong lensing clusters (cyan points and dashed lines) while simulations suggest $c$ is between 3.6 and 5.1 \citep[e.g.,][green and magenta points/lines, respectively]{duffy,maccio}.}
\label{fig_inferFree}
\end{figure*}

Our inference for the mass-concentration relationship of the CASSOWARY lenses is shown in Figure \ref{fig_inferFree}. The prior on $a$ is uniform between $-2$ and 0, the prior on $b$ is uniform between 0 and 2, and for the scatter we use a prior that is uniform in log $\sigma$ between 0.001 and 1. Our choice of the form of the prior for the halo masses (i.e., uniform or normal) does not significantly affect our results, although the normal prior leads to marginally more precise inference and smaller intrinsic scatter in the mass-concentration relation. We find that we are unable to constrain the slope $a$, although the data show a preference for a shallow slope. There is a significant tail towards steeper slopes that is covariant with the intercept such that the concentration of a halo with the sample mean mass ($M_{200} = 10^{14.2}~M_\odot$) is always $\sim7$. We robustly constrain the intercept to be $b = 0.85\pm0.13$, in spite of our poor inference on the slope; this is due to the limited dynamic range of halo masses, which we empirically find to span only $0.5-1$ dex ($\Delta M_{\rm U} \sim \sigma_{\rm N} \sim 0.2 - 0.5$).

\begin{figure*}
\centering
\includegraphics[width=0.96\textwidth,clip]{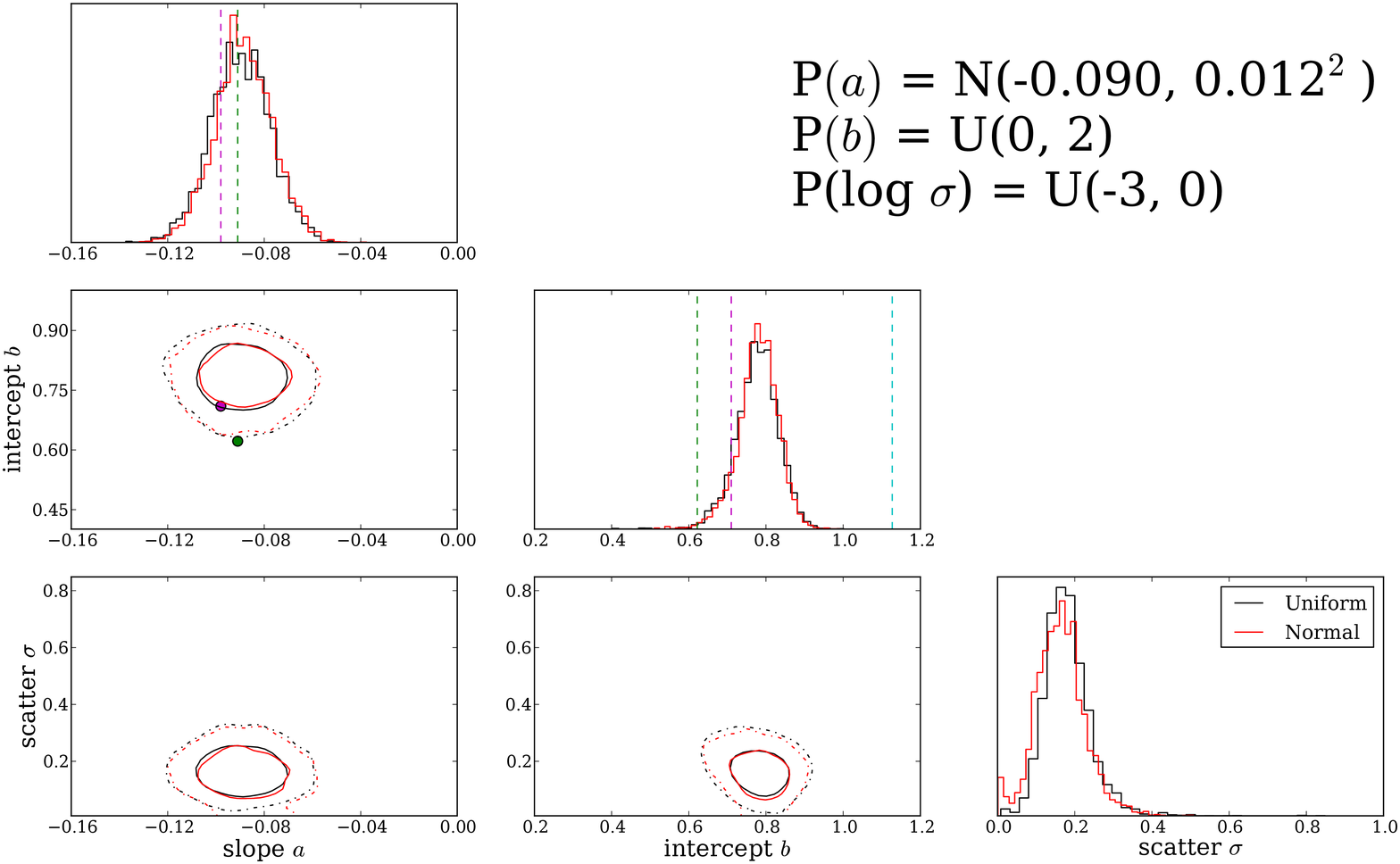}
\caption{The same inference as Figure \ref{fig_inferFree} but with a normal-distribution prior on the slope $a$ as motivated by simulations. The prior leads to slightly smaller inferred concentrations (log $c = 0.78\pm0.05$, $c \sim 6$, for a halo with mass $M_{200} = 10^{14}~M_\odot$) that agree with simulations at the 2-$\sigma$ level.}
\label{fig_inferDuff}
\end{figure*}

The small range of $M_{200}$ in our sample motivates us to also investigate a model with a prior on the slope consistent with the mass-concentration relations found in simulations (Figure \ref{fig_inferDuff}). We choose P($a$) to be normally distributed with mean 0.09 and variance $0.012^2$, consistent with, e.g., \citet{duffy} and \citet{maccio}. The imposition of this prior leads to significantly more precise inference on the intercept, which we find to be $b = 0.78\pm0.05$. Using a mean redshift of $z = 0.4$, this concentration is in good agreement with \citet{maccio}, who find the equivalent of $b = 0.71$, but is somewhat larger than the values of $0.56-0.62$ implied by \citet{duffy}. We find that the intrinsic scatter is $\sigma = 0.17\pm0.06$, also in agreement with simulated dark matter halos. We note that if we explicitly set $\Upsilon$ to a Salpeter or Chabrier value and repeat the inference, we find log~$c$ is between 0.64 and 0.82, respectively.

\section{Discussion}
We find that dark matter halos of groups and poor clusters (i.e., objects with log $M_{200} \sim 14$) appear to be only $17-66\%$ more concentrated than simulated dark matter halos of the same mass. This result appears to conflict with several recent analyses of strong lensing clusters that find systematically larger concentrations than simulations at lower cluster masses \citep[e.g.,][]{oguri12,wiesner}. Indeed, both of those results included strong lens systems investigated in this paper, so the differing conclusions are a bit alarming. However, there are several points that may resolve the tension between our results and these other observational findings.

\citet{wiesner} infer a steep slope but with a very large uncertainty that makes their measurement consistent with simulations at the 1.5-$\sigma$ level. Unfortunately they do not report an uncertainty on the intercept, nor do they provide details of the assumed prior on the slope or whether the posterior is Gaussian or highly skewed towards shallow values (e.g., Figure \ref{fig_inferFree}). Moreover, and perhaps most importantly, details of the fitting procedure are not provided, including whether the significant covariance between $c$ and $M_{200}$ is taken into account (e.g., Figure \ref{fig_inference}) and what prior is assumed for the distribution of halo masses.

If the prior on the halo masses (i.e., the contribution from the halo mass function and the CASSOWARY selection function) is assumed to be uniform and wide, then individual lens systems are given significant freedom to scatter along the degeneracy line illustrated by the black contours in Figure \ref{fig_inference}. These typically have slopes of $\sim -0.5$, consistent with the slopes found in previous strong lensing analyses \citep{oguri12,wiesner}. This suggests that, at least for the case of \citet{wiesner} who use the same type of constraints as we present here, these results may arise from fitting the observational scatter rather than the intrinsic mass-concentration relation.

The limited dynamic range of the CASSOWARY lenses presented here does not allow us to place interesting constraints on the slope of the mass-concentration relation, although our data are consistent with the slopes typically found in simulations (e.g., $a \sim -0.1$). We nevertheless find evidence for a slightly higher normalisation than dark-matter-only simulations \citep[e.g.,][]{duffy,maccio,neto}. \citet{oguri12} model the lensing selection function and find that this leads to a steeper inference on the mass-concentration relation and a higher normalisation than dark matter simulations predict for $M \approx 10^{14} M_\odot$. We do not find significant evidence for this over-concentration in our sample, although we explicitly fit (and subsequently marginalise over) the form of the parent population of the halos; this may signal that group-scale halos do not, in fact, suffer from an over-concentration problem.

\citet{gralla} find that, if the mass-concentration relation from simulations is assumed, their halo masses derived from Sunyaev-Zel'dovich effect measurements in strong lensing clusters predict Einstein radii $\approx 30$ per cent smaller than the observed Einstein radii. This result appears to disagree with what we find, but because it is not cast in terms of the slope and normalisation of the mass-concentration relation it is difficult to compare with our analysis. \citet{fedeli} has re-analysed several compilations of mass-concentration data from a variety of different observations and also infers a significantly steeper slope than simulations predict. He subsequently finds that halos can fit approximately the data if very efficient star formation and halo contraction due to baryonic cooling is included in his model. However, this model is difficult to reconcile with our result, and the huge discrepancy between the derived mass-concentration relations of \citet{fedeli} and e.g., \citet{ettori} indicates that his model may be fitting scatter in the data rather than intrinsic signal. On the other hand, Deason et al.~(2013) use strong lensing, a central stellar velocity dispersion profile, and satellite-galaxy kinematics to simultaneously constrain the normalisation of the central galaxy mass-to-light ratio and the group dark matter halo of the lens SDSSJ2158+0257, finding a halo with mass $M_{200} = 10^{14.2} M_\odot$ and concentration $c = 4.4$; this concentration is even \textit{lower} than what we find and agrees well with simulations.

\section{Conclusions}
We have investigated constraints on the mass-concentration relation of 26 massive group/poor cluster strong lenses by combining strong lens Einstein radius measurements with estimates of halo masses from a mass-richness relation. Contrary to previous studies, our lenses -- with typical mass $M_{200} = 10^{14.2}~M_\odot$ -- are only slightly ($\sim$ 0.1 dex) over-concentrated compared to halos in dark matter simulations, and we also find that we are unable to constrain the slope of the mass-concentration relation. This latter point is primarily due to the small range of halo masses spanned by our sample and the lack of precise constraints on the halo masses and concentrations for individual systems. Our halo mass dynamic range is inferred directly from the data by explicitly fitting for the mass selection function of the lenses, and we find that this is approximately two times smaller than the range of masses we would have found by considering, e.g., the distribution of the maximum likelihood masses for each system. We therefore conclude that explicitly modelling the halo mass selection function and directly accounting for the covariance between the `measured' halo masses and concentrations are required to make robust, unbiased inference.

Future prospects for constraining the slope of the mass-concentration relation are favourable. We have only presented constraints from a quarter of the lenses in the CASSOWARY sample, and our data can be combined with data from more massive clusters to improve our halo mass dynamic range and therefore get a longer lever arm for measuring the $M_{200}-c$ slope. Furthermore, two additional mass tracers can readily be obtained for the systems investigated here. Central stellar velocity dispersion measurements should help to significantly improve our constraints for individual systems (e.g., Deason et al.~2013), while measuring the radial magnifications of the lensed arcs allows the logarithmic slope of the projected total mass distribution to be constrained with $\sim 2\%$ precision at the Einstein radius \citep{dye}. Indeed, this latter measurement will also allow a direct investigation of one of the most significant assumptions in this paper, that the halos follow NFW profiles!


\section*{Acknowledgements}
We thank the referee for providing helpful comments, suggestions, and corrections. JMB acknowledges the award of a STFC research studentship. VB acknowledges financial support from the Royal Society. This work has made extensive use of the SDSS database. Funding for the SDSS and SDSS-II has been provided by the Alfred P. Sloan Foundation, the Participating Institutions, the National Science Foundation, the US Department of Energy, the National Aeronautics and Space Administration, the Japanese Monbukagakusho, the Max Planck Society, and the Higher Education Funding Council for England.

\label{lastpage}


\appendix
\section{Calculating Model Observables}\label{appendix}
The mass-richness relation described in Section 2.2 yields an estimate of $M_{500}$, the mass within $r_{500}$, where $r_{500}$ is defined as the radius at which the mean density enclosed by the mass distribution is 500 times greater than the critical density of the universe, $\rho_{\rm c}$. In order to compare this mass with our models we must compute $r_{500}$ for each set of model parameters $M_{200}$ and $c$. We note that the mean mass density for an NFW profile within radius $x$ (in units of the scale radius for that profile, $r_{\rm s} = r_{200}/c$) is given by
\begin{equation}
\overline \rho(x) = 3 \rho_0 x^{-3} \left[\mathrm{ln}\left(1+x\right) - \frac{x}{1+x}\right],
\label{eq:rhobar}
\end{equation}
where $\rho_0$ fixes the normalisation of the profile \citep[see, e.g.,][]{wbNFW}. The definition of $r_{500}$ implies that $\overline \rho(r_{500}/r_{\rm s}) = 500 \rho_{\rm c}$, and we therefore solve for $r_{500}/r_{\rm s}$ by approximating the inversion of Equation \ref{eq:rhobar} with a spline. Finally, the mass of the NFW profile within $r_{500}$ is given by,
$$
M_{500} = \overline \rho(r_{500}) \frac{4}{3} \pi r^3_{500} = 500 \rho_c  \frac{4}{3} \pi r^3_{500}.
$$
In principle, the model $M_{500}$ is then the sum of this and the baryonic mass within $r_{500}$, but we neglect the effect of baryons on $M_{500}$ because the baryon fraction within $r_{500}$ is typically quite small \citep[i.e., less than 10 per cent;][]{ettori}.

The lensing data offer two possibilities for comparing with the model: we may choose to determine the model Einstein radius and compare with the observed Einstein radius, or we may calculate the mass of the model within the observed Einstein radius and compare with the measured mass from our observations. We have chosen to employ the former option because the latter requires the assumption that we know the aperture (the Einstein radius) with perfect precision. The Einstein radius is defined as the radius within which the mean surface mass density is equal to the lensing critical density,
$$
\overline \Sigma(r_{\rm Ein}) = \Sigma_{\rm crit}
$$
where
$$
\Sigma_{\rm crit} = \frac{c^2 D_{\rm s}}{4\pi G D_{\rm l}D_{\rm ls}}
$$
with $D_{\rm s}, D_{\rm l}$, and $D_{\rm ls}$ the angular diameter distances to the source, lens, and between the lens and source, respectively. In practice, our mass density comes from the dark matter component and the baryonic component, and therefore
$$
\overline \Sigma_{\rm NFW}(r_{\rm Ein}) + \overline \Sigma_*(r_{\rm Ein}) = \Sigma_{\rm crit}.
$$
The mean mass surface density for an NFW halo is
\begin{equation}
\overline \Sigma_{\rm NFW}(x) = r_{\rm s} \rho_0 \mathcal{F}(x)
\label{eq:signfw}
\end{equation}
with $\mathcal{F}(x)$ provided by \citet{wbNFW}. We assume that $\overline \Sigma_*(r_{\rm Ein})$ is only contributed to by the luminosity of the lensing galaxy and that all of the lens galaxy light, $L_{\rm lens}$, is within $\widetilde r_{\rm Ein}$; for simplicity, here we also ignore the uncertainty on $\widetilde r_{\rm Ein}$ so that the only free parameter for this component is the stellar mass-to-light ratio, i.e., 
$$
\overline \Sigma_*(r_{\rm Ein}) = \frac{\Upsilon L_{\rm lens}}{\pi \widetilde r^2_{\rm Ein}}.
$$
The Einstein radius is therefore given by
\begin{equation}
\Sigma_{\rm crit} - \frac{\Upsilon L_{\rm lens}}{\pi \widetilde r^2_{\rm Ein}} = r_{\rm s} \rho_0 \mathcal{F}(r_{\rm Ein}/r_{\rm s}),
\label{eq:solveRein}
\end{equation}
which can be solved by approximating the inversion of $\mathcal{F}$ with a spline.

\onecolumn

\section{Hierarchical Inference Statistical Model}\label{appendixb}
The typical scheme for fitting a linear relation with scatter includes defining `true' parameters for the independent and dependent variables and subsequently marginalising over these parameters \citep[e.g.,][]{kelly}; this is usually only done implicitly, as the marginalisation often can be performed analytically in the case of Gaussian uncertainties on the measured parameters. In our analysis, an explicit definition of the variables is necessary (e.g., Equation 1) because \textit{we are not fitting a line to our measurements} $\widetilde r_{\rm Ein}$ and $\widetilde M_{500}$ (indeed, even these are not really `measurements' but are instead quantities derived from the SDSS imaging). 

Our full inference for a single lens can be written as
\begin{eqnarray}
\nonumber \mathrm{P}(a,b,\sigma|\widetilde r_{\rm Ein},\widetilde M_{500}) & = & \int \mathrm{P}(\widetilde r_{\rm Ein},\widetilde M_{500}|r_{\rm Ein}(c,M_{200},\Upsilon),M_{500}(c,M_{200}))\ {\rm P}(c,M_{200},\Upsilon|a,b,\sigma,{\rm log}\ M_{\rm X},\Delta M) \\
& & {\rm P}(a,b,\sigma,{\rm log}\ M_{\rm X},\Delta M)\ {\rm d}c\ {\rm d}M_{200}\ {\rm d}\Upsilon\ {\rm d}{\rm log}\ M_{\rm X}\ {\rm d}\Delta M
\end{eqnarray}
where we use the compact notation ${\rm log}\ M_{\rm X}$ and $\Delta M$ to denote the mean and width of the priors on $M_{200}$ for the uniform and normal distribution models, as described in Section 4.1. The first term on the right is, suppressing the explicit dependence on $c$, $M_{200}$, and $\Upsilon$,
\begin{eqnarray}
\nonumber {\rm P}(\widetilde r_{\rm Ein},\widetilde M_{500}|r_{\rm Ein},M_{500}) & = & {\rm P}(\widetilde r_{\rm Ein}|r_{\rm Ein})\ {\rm P}(\widetilde M_{500}|M_{500}) \\
 & = & \frac{1}{2\pi \sigma_{\rm Ein} \sigma_{M}}e^{-\frac{1}{2}\frac{(\widetilde r_{\rm Ein} - r_{\rm Ein})^2}{\sigma^2_{\rm Ein}}}e^{-\frac{1}{2}\frac{(\widetilde M_{\rm 500} - M_{\rm 500})^2}{\sigma^2_{M}}},
\end{eqnarray}
with $\sigma_M$ the uncertainty on $\widetilde M_{500}$. Furthermore, following Equations 3 and 4,
\begin{eqnarray}
{\rm P}(c,M_{200},\Upsilon|a,b,\sigma,{\rm log}\ M_{\rm X},\Delta M) & = & {\rm P}({\rm log}\ c|a,b,\sigma,{\rm log} M_{200})\ {\rm P}({\rm log}\ M_{200}|{\rm log}\ M_{\rm X},\Delta M) {\rm P}(\Upsilon)
\end{eqnarray}
with
\begin{eqnarray}
{\rm P}({\rm log}\ c|a,b,\sigma,{\rm log} M_{200}) & \sim & {\rm N}(a~{\rm log}\  M_{200}/10^{14} + b,\sigma^2) \\
{\rm P}({\rm log}\ M_{200}|{\rm log}\ M_{\rm X},\Delta M) & \sim & \begin{cases}
{\rm U}({\rm log}\ M_{\rm U}-\Delta M_{\rm U},{\rm log}\ M_{\rm U}+\Delta M_{\rm U}) \\
{\rm N}({\rm log}\ M_{\rm N},\sigma^2_{\rm N}) \end{cases} \\
{\rm P}(\Upsilon) & \sim & {\rm U}(1.8,3.2).
\end{eqnarray}
Finally, our priors on the hyper-parameters, the final term in Equation B1, are all independent and are
\begin{eqnarray}
{\rm P}(a) & \sim & \begin{cases} {\rm U}(-2,0) \\ {\rm N}(-0.090,0.012^2) \end{cases} \\
{\rm P}(b) & \sim & {\rm U}(0,2) \\
{\rm P}({\rm log}\ \sigma) & \sim & {\rm U}(-3,0) \\
{\rm P}({\rm log} M_{\rm X}) & \sim & {\rm N}(14,1) \\
{\rm P}({\rm log}\ \Delta M) & \sim & {\rm U}(-1.7,0.3).
\end{eqnarray}

It may appear that Equation B1 is just an integral over a product of Gaussian and uniform distributions and should therefore be analytically tractable. However, Equation B2 is only Gaussian in the transformed variables $r_{\rm Ein}$ and $M_{500}$ and it actually has a non-trivial form in the variables of integration $c$, $M_{200}$, and $\Upsilon$ (e.g., Figure 1). In practice, our inference is performed by first evaluating Equation B2 for each lens on a grid of $c$, $M_{200}$, and $\Upsilon$. The $\Upsilon$ marginalisation integral can then be evaluated, and we use a Markov chain Monte Carlo simulation to generate samples for $a$, $b$, $\sigma$, ${\rm log}\ M_{\rm X}$, and $\Delta M$ as the solution to Equation B1.

\end{document}